\documentclass[twocolumn]{aastex631}

\usepackage{xcolor}
\usepackage{hyperref}

\newcommand{\hess}{HESS J1303-631}
\newcommand{\psr}{PSR J1301-6305}
\newcommand{\xmm}{\textit{XMM-Newton}}
\newcommand{\chandra}{\textit{Chandra}}
\newcommand{\fermi}{\textit{Fermi}-LAT}
\newcommand{\nh}{N_{\rm H}}
\newcommand{\fx}{F_{2-10\,{\rm keV}}}

\shorttitle{Deep X-ray Observations of \hess}
\shortauthors{Gagnon et al.}
\submitjournal{ApJ}

\begin{document}

\title{Deep X-ray Observations of the Gamma-ray Bright Pulsar Wind Nebula \hess}

\author[0000-0003-0902-1935]{Seth Gagnon}
\affiliation{Department of Physics, The George Washington University, 725 21st St, NW, Washington, DC 20052, USA}

\author[0000-0002-6447-4251]{Oleg Kargaltsev}
\affiliation{Department of Physics, The George Washington University, 725 21st St, NW, Washington, DC 20052, USA}

\author[0009-0009-1950-6937]{Yichao Lin}
\affiliation{Department of Physics, The George Washington University, 725 21st St, NW, Washington, DC 20052, USA}

\author[0000-0002-8548-482X]{Jeremy Hare}
\affiliation{Astrophysics Science Division, NASA Goddard Space Flight Center, Greenbelt, MD, 20771, USA}
\affiliation{Center for Research and Exploration in Space Science and Technology, NASA/GSFC, Greenbelt, Maryland 20771, USA}
\affiliation{The Catholic University of America, 620 Michigan Ave., N.E. Washington, DC 20064, USA}

\author[0000-0001-9633-3165]{Jordan Eagle}
\affiliation{Astrophysics Science Division, NASA Goddard Space Flight Center, Greenbelt, MD, 20771, USA}

\correspondingauthor{Seth Gagnon}
\email{sethgagnon@gwmail.gwu.edu}

\begin{abstract}
HESS J1303$-$631 is an evolved, TeV-bright pulsar wind nebula (PWN) powered by the
energetic radio pulsar PSR~J1301$-$6305, offset by $\approx7'$ northwest of the TeV
centroid, with an energy-dependent gamma-ray morphology. We present new, deep joint
\textit{Chandra} (108\,ks) and \textit{XMM-Newton} (102\,ks) observations of this field.
High-resolution \textit{Chandra} ACIS imaging resolves the pulsar from a neighboring
point source that \textit{XMM-Newton} cannot separate. The pulsar's 2--10\,keV spectrum
is well described by an absorbed power law  with $\Gamma=2.02\pm1.04$ and an unabsorbed luminosity of
$3.44\times10^{31}\,{\rm erg\,s^{-1}}$ at $d=4.7$\,kpc. \textit{Chandra} also reveals
faint diffuse emission east of the pulsar: a clumpy compact nebula on arcsecond scales
($\Gamma=2.98\pm0.53$, $L_X=4.15\times10^{31}\,{\rm erg\,s^{-1}}$) and a fainter extended
nebula on arcminute scales ($\Gamma=1.20\pm1.25$, $L=4.36\times10^{31}\,{\rm erg\,s^{-1}}$).
This morphology, the hard GeV counterpart, and the energy-dependent TeV emission together
support an evolved, synchrotron-underluminous PWN whose X-ray nebula is far more compact
than its TeV nebula. We also serendipitously discovered a transient X-ray source,
CXOU~J130131.0$-$625919, in one of the nine new \textit{Chandra} observations. Its hard
spectrum ($\Gamma=1.4\pm0.3$) and 0.5--7\,keV luminosity of
$3.34\times10^{32}\,{\rm erg\,s^{-1}}$ at a fiducial distance of 3\,kpc, together with a
plausible near-infrared counterpart, most closely match a flare from a young pre-main-sequence
star, though an active binary or compact-object binary is not excluded. Its large offset
from HESS J1303$-$631 makes association with the TeV source unlikely.
\end{abstract}


\section{Introduction} \label{sec:intro}

Pulsar wind nebulae (PWNe) are efficient accelerators of relativistic electrons and positrons, and
they are now one of the dominant identified classes of Galactic TeV sources
\citep[e.g.,][]{gaensler2006,hessgps2018}.  Their multiwavelength appearance depends strongly on
age and environment.  Synchrotron-emitting X-ray particles cool rapidly and remain concentrated near
the pulsar, whereas inverse-Compton-emitting particles can survive longer and fill much larger GeV
and TeV emitting nebulae.  In evolved systems, reverse-shock interaction with the host supernova remnant and
pulsar proper motion can displace the relic particle population from the currently powered compact
X-ray nebula \citep[e.g.,][]{gelfand2009}.

In the crowded regions of the Galactic plane many TeV sources remain unidentified or ambiguously identified due to multiple possible (or a complete lack of) contributors of energetic particles. The known types of Galactic accelerators, in addition to pulsars with PWNe, include microquasars, high-mass gamma-ray binaries, supernova remnants, and, possibly, young stellar objects \citep[e.g.,][]{2009ARA&A..47..523H}. On some occasions an AGN of a blazar type can be seen as a TeV source through the Galactic plane \citep[e.g.,][]{2011A&A...529A..49H, 2018ApJ...862...41A}. To further complicate matters, in many cases X-ray counterparts of these accelerators are faint or relatively faint in X-rays. Therefore, a wholistic analysis and assessment of multiple sources is often needed to make a judgment about the nature of the TeV source. 

\hess\ was discovered with H.E.S.S. \citep{2005A&A...439.1013A} as a bright, extended
TeV source in the Galactic plane ($l = 304.2^\circ, b = -0.3^\circ$)
with and integrated flux above 380 GeV of
$(1.2\pm0.2)\times 10^{-11}$ cm$^{-2}$ s$^{-1}$, about 17\% of the Crab Nebula flux, and a TeV
photon index $\Gamma=2.44\pm0.05_{\rm stat}\pm0.2_{\rm syst}$.  A short ($\approx 5$ ks) archival \chandra\ ACIS-I
observation did not reveal an obvious X-ray counterpart and
only placed a limit on pulsar and PWN emission
and diffuse X-ray emission \citep{mukherjee2005}.  This
made the source one of the early
``dark accelerator'' candidates.

The source was later 
studied in more detail by the H.E.S.S. Collaboration
\citep{hess2012} who found that the morphology is well described by an asymmetric Gaussian with
$\sigma_x=0.194\pm0.008^\circ$, $\sigma_y=0.145\pm0.006^\circ$, and position angle
$147\pm6^\circ$.  The large-region TeV spectrum is not well represented by a simple power law, with a
better fit found for a
cutoff power-law (PL) model having $\Gamma=1.5\pm0.2$ and cutoff energy
$E_{\rm cut}=7.7\pm2.2$ TeV \citep{hess2012}. A young (spin-down age 11 kyrs), energetic ($\dot{E}=1.7\times 10^{36}$ erg s$^{-1}$) \psr\ is located  $\approx7'$  NW of the TeV source center. It has been proposed as a source of energetic particles powering the TeV emission. Additional support for the evolved PWN scenario comes from 
the energy-dependent  morphology of the TeV source: lower-energy
0.84--2 TeV photons form a broad, displaced nebula, while the $>10$ TeV emission is more compact and
closer to the pulsar. The distance to the pulsar inferred from its dispersion measure (DM=374 cm$^{-3}$ pc) varies greatly depending on the galactic electron density distribution model used: 10.7 kpc (YMW17; \citealt{2017ApJ...835...29Y}), 6.6 kpc (cl02; \citealt{2002astro.ph..7156C}), and 4.7 kpc (OC26; \citealt{2026ApJ..1002....3O}). We assume the latter value for the rest for the paper. 

\cite{hess2012} also reported a faint \xmm\ counterpart associated with \psr.  The X-ray emission consists
of a compact source near the pulsar plus a diffuse component extending roughly eastward toward the
TeV nebula.  The diffuse feature was detected at $\sim6.5\sigma$ in the archival \xmm\ data and was
modeled with an absorbed PL with   poorly constrained
$\nh=2.7^{+1.3}_{-1.1}\times10^{22}$ cm$^{-2}$,
$\Gamma_X=2.0^{+0.6}_{-0.7}$, and unabsorbed
$\fx=1.6^{+0.2}_{-0.4}\times10^{-13}$ erg cm$^{-2}$ s$^{-1}$ \citep{hess2012}.
The X-ray nebula is much smaller than the TeV nebula, as expected for an evolved PWN in which the
highest-energy synchrotron-emitting electrons cool rapidly.  The one-zone leptonic model presented
by \citet{hess2012} 
obtained a magnetic field of $1.4\,\mu$G, but the authors noted that the
very different X-ray and TeV sizes
require a more complex treatment.

At GeV energies, \fermi\ analyses detected a hard, extended counterpart 4FGL J1303.0-6312e.  The 14 yr \fermi\ study by
\citet{zhou2023} measured a 10--500 GeV photon index of $1.69\pm0.09$ and a 2D Gaussian extension
$R_{68}=0.32\pm0.04^\circ$, strengthening the evolved-PWN interpretation.

Dedicated ATCA observations did not detect an extended radio
PWN  at 5.5 or 7.5 GHz, and archival 1.384 and 2.368 GHz data also did not show a significant radio
counterpart to the X-ray nebula \citep{sushch2015}.  However, 
ATCA data revealed a 
shell-like radio
structure (likely a supernova remnant)  $15'$ east of the pulsar, but its association with \psr\ is questionable because it would require a very large pulsar velocity, $1,800$ km s$^{-1}$ for the smallest $d=4.7$ kpc and pulsar age of 11 kyrs. Even if the true pulsar's age is a factor of two larger, the velocity would still be high enough to expect a formation of a strong bowshock around a high-Mach pulsar. 

In order to better resolve the compact PWN from the pulsar additional Chandra observations (PI J.~Eagle; ObsIDs 28056, 28509, 28510, 29079, 29080, 29192, 29204, 29205, 29206) were carried out. It was accompanied by an XMM-Newton observation (PI J.~Eagle; ObsID 0931990101) to collect more photons for spectral analysis. In Section \ref{sec:observations} we provide details of these observations followed by the data analysis results in Section \ref{sec:analysis}. We discuss the results and provide multiwavelength context (including new imaging form MeerKAT GPS; \citealt{2024MNRAS.531..649G}) in Section \ref{sec:discussion}. We conclude with a summary in Section \ref{sec:summary}. 

\section{Observations} \label{sec:observations}

Ten recent Chandra ACIS observations of \psr\ region  are summarized in Table \ref{tab:obs}. 
Nine of these were taken between Nov 2023 and Jan 2024. The target was imaged on ACIS S3 detector near the aim-point close to optical axis of the telescope.
One observation was taken in 2004 (ObsID: 6098), and is not included in the analysis because of the short exposure time (4.7 ks)
and \psr\ being imaged $\sim9'$ off-axis.
It is possible that this exposure could have collected a comparable number of photons as the recent observations since the effective area at lower energies was much larger in the past.
However, there are very few counts at the location of the pulsar, suggesting a large NH or very weak/absent thermal component. 

XMM-Newton EPIC observed the \hess\ region on 3 occasions (see Table \ref{tab:obs}). The deepest observation aimed at \psr\  was carried out on  August 31, 2024.
Table~\ref{tab:obslog} summarizes X-ray observations of the \hess\  field. 

\begin{deluxetable*}{ccccl}
\label{tab:obslog}
\tablecaption{Chandra and XMM-Newton Observations of HESS J1303$-$631}
\tablewidth{0pt}
\tablehead{
\colhead{ObsID} & \colhead{Instrument} & \colhead{Exposure (ks)} & \colhead{Observation Date} & \colhead{Notes} 
}
\startdata
\multicolumn{5}{l}{\textit{Chandra ACIS}} \\
6098  & ACIS-I & 4.65  & 2004 Sep 25 & Not included in merged images or spectral analysis \\
28056 & ACIS-S & 9.94  & 2024 Jan 16 & \\
28509 & ACIS-S & 10.93 & 2023 Nov 26 & \\
28510 & ACIS-S & 10.12 & 2024 Jan 08 & \\
29079 & ACIS-S & 10.93 & 2024 Jan 12 & \\
29080 & ACIS-S & 15.87 & 2024 Dec 02 & \\
29192 & ACIS-S & 11.62 & 2024 Jan 12 & \\
29204 & ACIS-S & 11.52 & 2024 Jan 16 & \\
29205 & ACIS-S & 11.80 & 2024 Jan 16 & \\
29206 & ACIS-S & 10.64 & 2024 Jan 18 & \\
\hline
Total (ACIS) & \nodata & 108.02 & \nodata & \nodata \\
\hline 
\multicolumn{5}{l}{\textit{XMM-Newton EPIC (M1/M2/PN)}} \\
0303440101 & EPIC & $\sim$30/30/28 & 2005 Jul 12 & Archival; affected by chip-gap placement for the PWN region \\
0302340101 & EPIC & $\sim$31/30/26 & 2005 Jul 14 & Archival; significantly offset \\
0931990101 & EPIC & $\sim$70/77/48 & 2024 Aug 31 & New; exposure limited by significant flaring \\
\hline
Total (EPIC) & \nodata & 131/137/102 & \nodata & \nodata \\
\enddata
\tablecomments{All Chandra observations were taken in Timed Exposure (TE) mode with VFAINT telemetry. ObsID 6098 was imaged $\sim9'$ off-axis and was not included in the merged images or spectral analysis; the 108.02\,ks ACIS total above includes it, while the images and spectra analyzed in this paper use only the nine 2023--2024 pointings (103.37\,ks). XMM-Newton exposures are effective, background-flare-filtered exposure times for MOS1, MOS2, and pn, respectively, in ks.}
\label{tab:obs}
\end{deluxetable*}

\section{Data Reduction and Analysis} \label{sec:analysis}

\subsection{Chandra imaging and spectroscopy of \psr\ and its PWN.}

\begin{figure}
\centering
\includegraphics[width=0.49\textwidth]{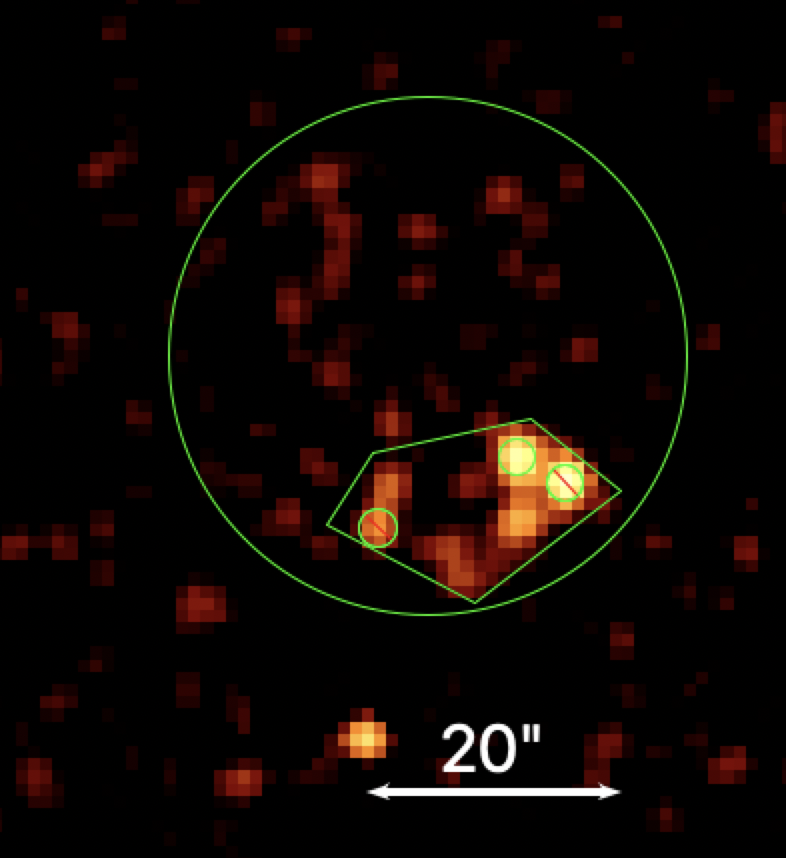}
\includegraphics[width=0.49\textwidth]{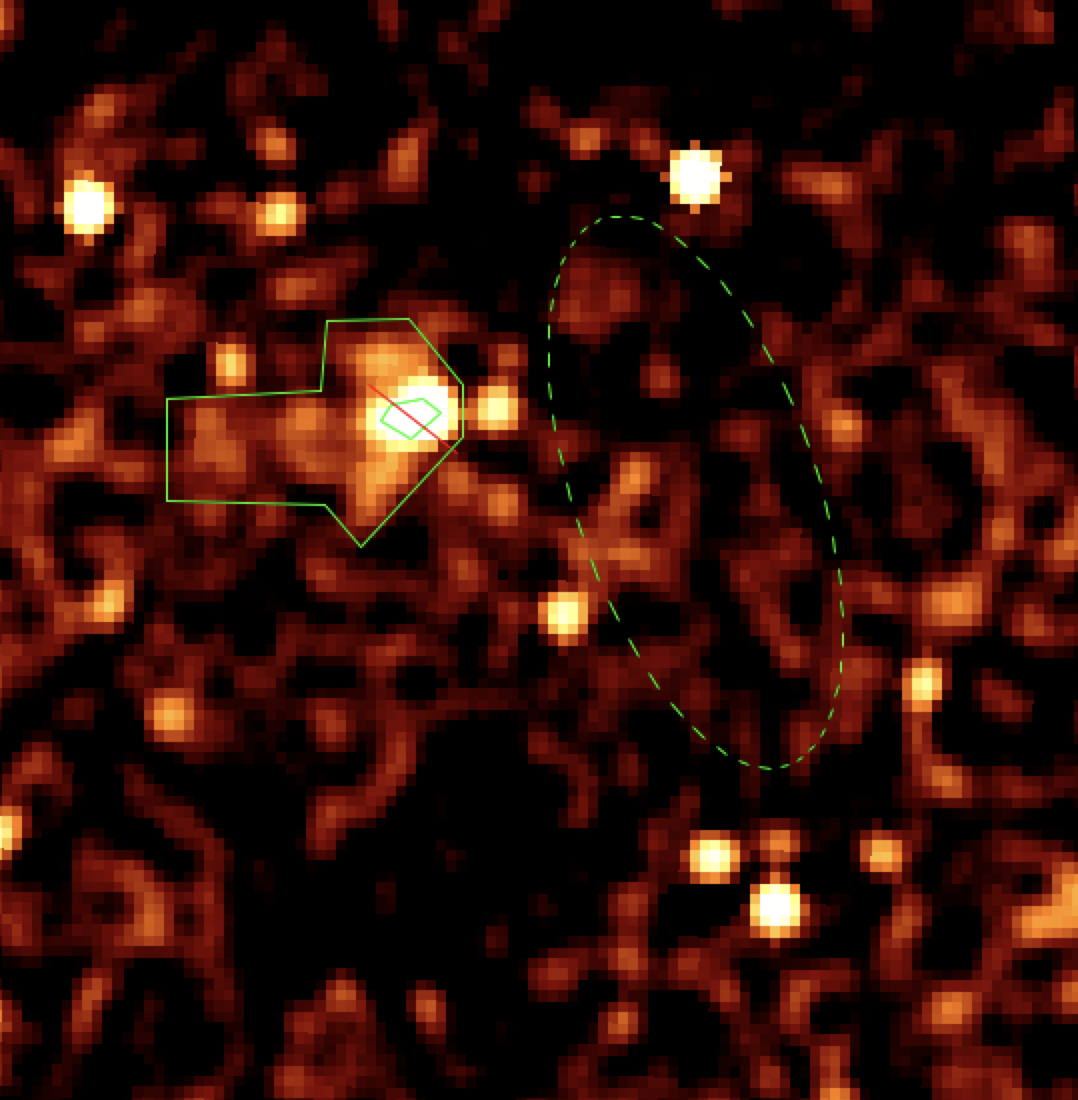}
\caption{Images of the 2--8 keV compact emission around PSR J1301--6305. Top: binned by a factor of 2 and smoothed (r=2, $\sigma$=1). The PSR (inner circle), Compact Nebula (polygon), and Extended Nebula (outer circle) regions used for spectral extraction are shown in green. Bottom: Exposure corrected image,  smoothed (r=2, $\sigma$=1). The Eastern Extension region (solid green polygon) with its associated background (dashed ellipse)used for spectral extraction is shown.}
\label{fig:cxo:spec:reg}
\end{figure}

\begin{figure}
    \centering
    \includegraphics[width=1.0\linewidth]{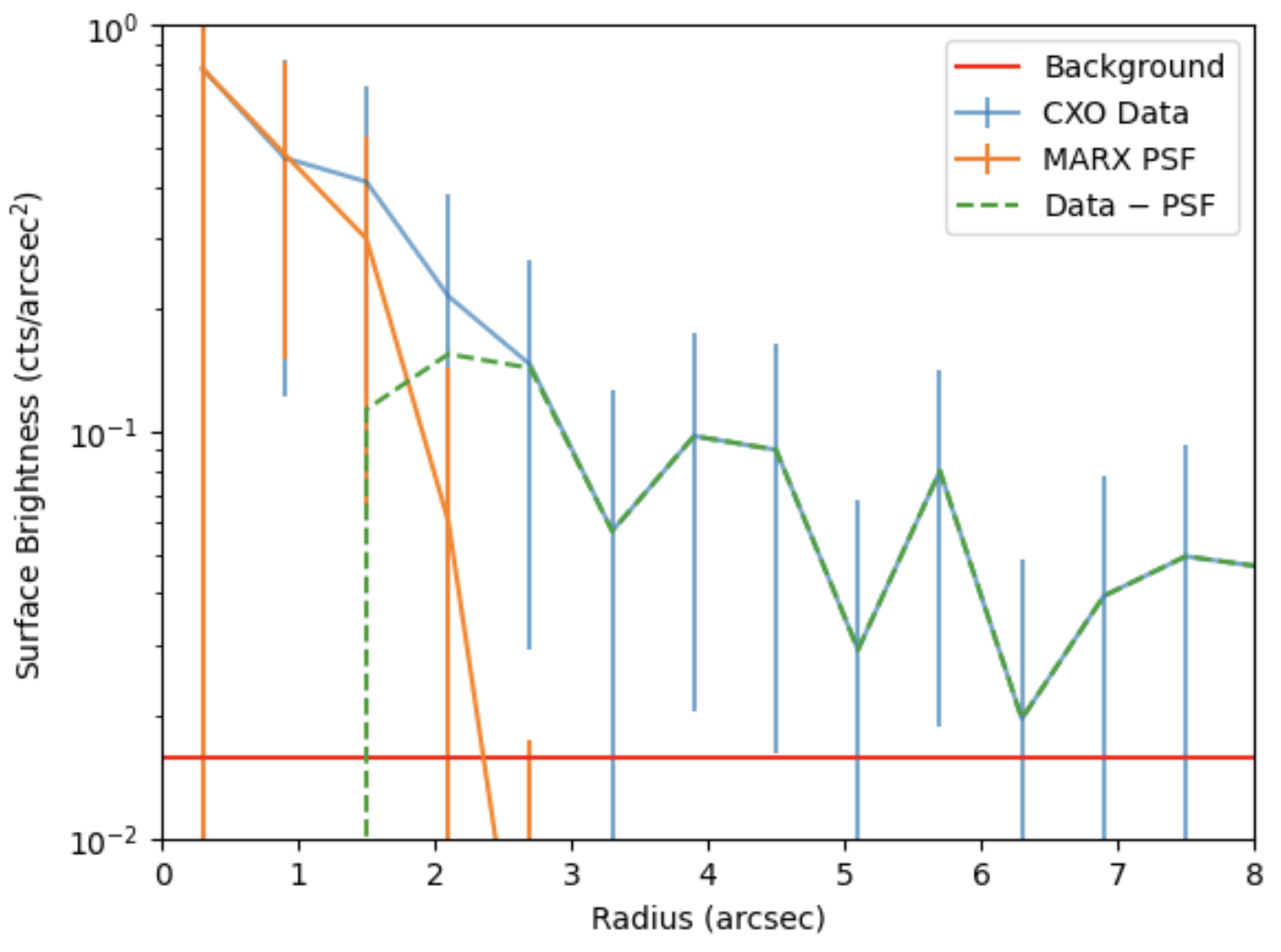}
    \caption{Radial profile of the J1301 PWN. The CXO data is shown in blue, the MARX PSF in orange and the difference between these two is shown by the dashed green line. The background level is shown by the red line.}
    \label{fig:cxo_rad_prof}
\end{figure}

The merged ACIS image of PSR J1301--6305 (hereafter J1301) is shown in Figure \ref{fig:cxo:spec:reg}. 
On small scales (upper panel in the figure) it appears that the compact nebula (polygon region) is actually composed of multiple point sources. 
There does appear to be excess emission above the background surrounding the pulsar, indicating a possible compact nebula extending roughly to the east of the pulsar, but the exact nature of the morphology is unclear given its faint X--ray flux, $F_{0.5-7}=(4.47\pm0.92)\times10^{-14}$ ergs s$^{-1}$ cm$^{-2}$.
To quantify the extension of this emission we compared the radial profile of the CXO data to that of a MARX point-source simulation (see Figure \ref{fig:cxo_rad_prof}). We see that there is significant extended emission  beyond the PSF ($\gtrsim 2''$).
Faint, diffuse emission is seen on larger scales (circular region, r=20$''$) extending to the NE of the pulsar. 
Unfortunately, even though we have a large baseline between the observations ($\sim$20 years) there are too few counts in the 2004 exposure to constrain the proper motion.

We performed a spectral analysis of J1301 and the surrounding extended emission.
The CXO observations included in the fit are summarized in Table \ref{tab:obs}. 
The older observation (ObsID: 6098) is not included due to the short exposure time and large off-axis angle. Therefore, astrometric correction is unnecessary.
The spectra were fit in XSPEC (v.4.15.0; \citealt{2001SPIE.4477...76F})
using their implementation of the Cash statistic ({\tt cstat}).

The model fit
to the pulsar and nebula
was XSPEC's absorbed power-law (PL) where the interstellar extinction is modeled with the {\tt tbabs} model (absorption cross sections and abundances described in \citealt{2000ApJ...542..914W}).
The spectra from all CXO observations were jointly fit in the 2--10 keV energy range.
The column density was frozen to $n_H=2.7\times10^{22}$ cm$^{-2}$, obtained  by \cite{2012A&A...548A..46H} from a fit to the PWN region using the archival XMM Newton observations.
The regions used for spectral extraction are shown in Figure \ref{fig:cxo:spec:reg}, and the results are summarized below and in Table \ref{tab:fits}.

\begin{deluxetable*}{lccccccc}
\tablecaption{Spectral Fit Results}
\tablewidth{0pt}
\tablehead{
\colhead{Region} &
\colhead{$N_{\rm H}$} &
\colhead{$\Gamma$} &
\colhead{Norm} &
\colhead{C-stat/dof} &
\colhead{$F_{2-10\,\mathrm{keV}}$} &
\colhead{$L_{2-10\,\mathrm{keV}}$}
}
\startdata
PSR & $2.7$\,(fixed) & $2.02 \pm 1.04$ & $(6.15 \pm 5.36)\times10^{-6}$ & 17.97 / 22 & $(1.30^{+0.20}_{-0.98})\times10^{-14}$ 
& $3.44\times10^{31}$ \\
Compact PWN & $2.7$\,(fixed) & $2.98 \pm 0.53$ & $(3.02 \pm 1.98)\times10^{-5}$ & 108.43 / 109 & $(1.57^{+0.14}_{-0.71})\times10^{-14}$ 
& $4.15\times10^{31}$ \\
Extended PWN & $2.7$\,(fixed) & $1.20 \pm 1.25$ & $(2.03 \pm 3.43)\times10^{-6}$ & 372.65 / 356 & $(1.65^{+0.12}_{-1.65})\times10^{-14}$ 
& $4.36\times10^{31}$ \\
\enddata
\tablecomments{
Spectra were fit with an absorbed power-law model using {\tt tbabs}. 
Column densities are in units of $10^{22}\ \mathrm{cm^{-2}}$. 
Fluxes are unabsorbed and reported in units of $\mathrm{erg\ s^{-1}\ cm^{-2}}$. 
Luminosities are unabsorbed and assume a distance of 4.7\,kpc.
Uncertainties are quoted at the 1$\sigma$ level.
}
\label{tab:fits}
\end{deluxetable*}

The pulsar region
has $30.00\pm5.66$ net counts after background subtraction, and the fit
yields a PL photon index of $\Gamma=2.02\pm1.04$, which is within the typical range seen for pulsars powering a PWN. The corresponding unabsorbed flux is $F_{2-10}=(1.30^{+0.20}_{-0.98})\times10^{-14}$ ergs s$^{-1}$ cm$^{-2}$
and the unabsorbed luminosity is $3.44\times10^{31}$ erg s$^{-1}$ for a distance of 4.7 kpc. 

The compact nebula with point sources excised
has $67.77\pm17.09$ net counts
and the spectrum 
is softer with $\Gamma=2.98 \pm 0.53$,
which is somewhat softer than typical values seen for PWN. However, this could be simply due to the large uncertainty in the measurement. 
The unabsorbed flux 
in 2-10 keV range is
$F_{2-10}=(1.57^{+0.14}_{-0.71})\times10^{-14}$ ergs s$^{-1}$ cm$^{-2}$ and an unabsorbed luminosity of $4.15\times10^{31}$ erg s$^{-1}$.

Finally, the extended nebula 
with the compact nebula excised
is much fainter with only $14.68\pm32.38$ counts after background subtraction
and the
spectrum  yields $\Gamma=1.20 \pm 1.25$, an unabsorbed flux of $F_{2-10}=(1.65^{+0.12}_{-1.65})\times10^{-14}$ ergs s$^{-1}$ cm$^{-2}$, and corresponding unabsorbed luminosity of $4.36\times10^{31}$ erg s$^{-1}$ at $d=4.7$ kpc. 

\begin{figure*}
\centering
\includegraphics[width=0.99\textwidth]{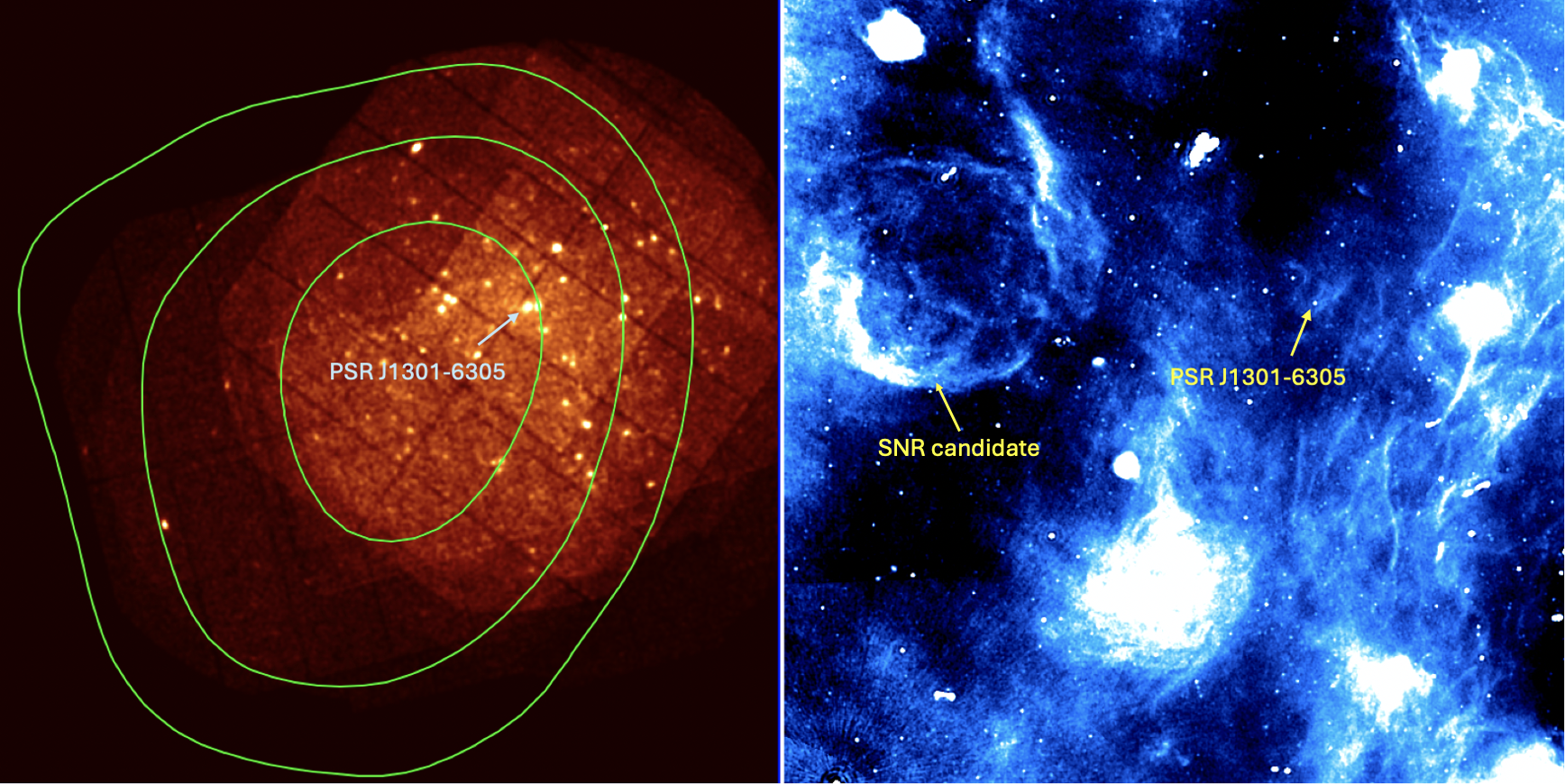}
\includegraphics[width=0.99\textwidth]{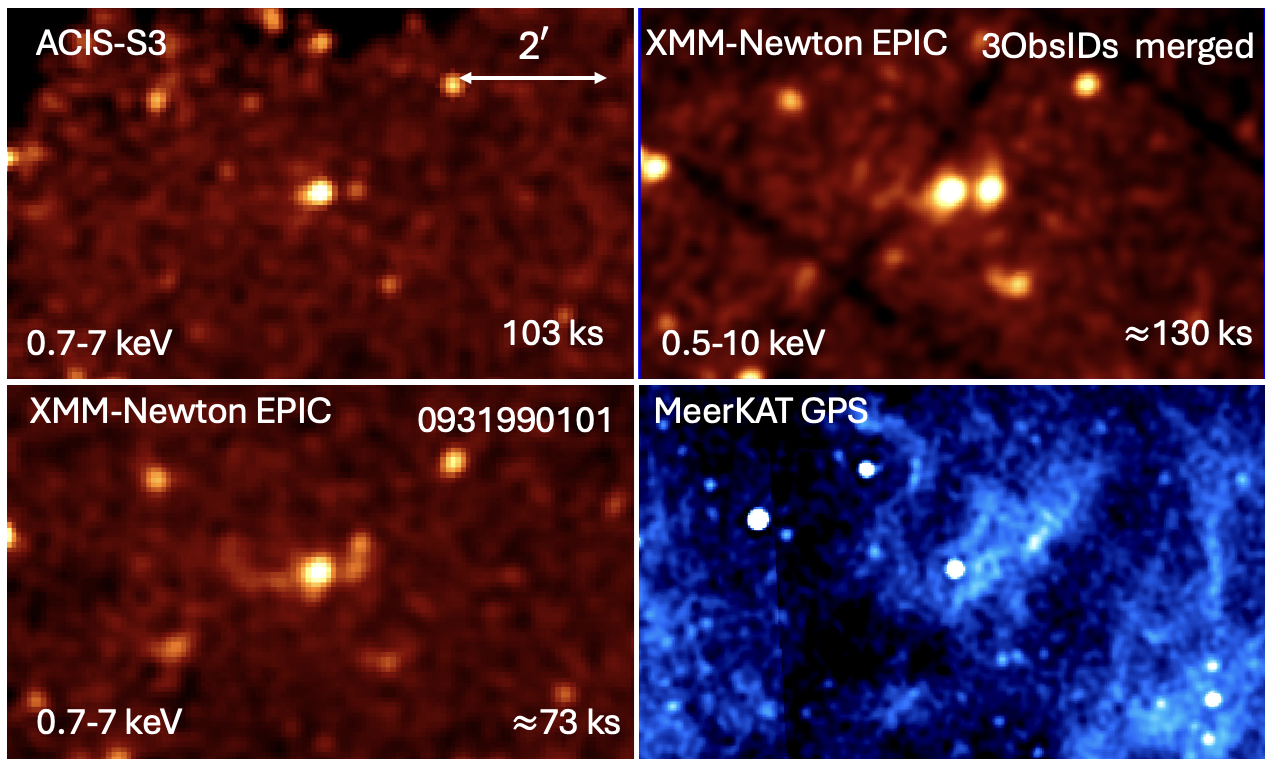}
\caption{Multiwavelength view of the field around PSR~J1301$-$6305 and
HESS~J1303$-$631. \textit{Top}: wide-field ($\sim30'$) \textit{XMM-Newton} EPIC
mosaic (0.5--10\,keV; left, red/orange) with H.E.S.S. TeV surface-brightness
contours overlaid in green, and the corresponding SARAO MeerKAT GPS 1.3\,GHz
radio image (right, blue) showing the shell-like SNR candidate $\sim15'$ east
of the pulsar and diffuse radio emission extending west of the pulsar;
PSR~J1301$-$6305 is marked with an arrow in both panels. \textit{Bottom two
rows}: comparison of the same region in
\textit{Chandra} ACIS-S3 (0.7--7\,keV, 103\,ks), the merged \textit{XMM-Newton}
EPIC image (0.5--10\,keV, three ObsIDs combined, $\approx130$\,ks), the new,
deep \textit{XMM-Newton} EPIC observation alone (ObsID 0931990101, 0.7--7\,keV,
$\approx73$\,ks), and the MeerKAT GPS radio image (bottom right), all shown at
the same field of view (2$'$ scale bar, top of the ACIS-S3 panel).}
\label{fig:xmm:mw}
\end{figure*}

\subsection{\xmm\   imaging and spectroscopy of \psr\ and its PWN.}

We used the {\tt images}\footnote{\url{https://www.cosmos.esa.int/web/xmm-newton/images}} script to reduce the XMM-Newton data and make a exposure corrected image of the field. The data were processed with SAS version 1.3 using the standard procedures. MOS1 and 2 events were selected with PATTERN $\le 12$ and pn events with
PATTERN $\le 4$, We removed background particle flares for the PN and MOS1/2 using a count rate cut of 0.4 cts s$^{-1}$ and 0.35 cts s$^{-1}$, respectively. The images were created in the 0.5-10 keV band for each of the three ObsIDs individually and then merged using the {\tt emosiac} SAS task. Figure~\ref{fig:xmm:mw} shows the merged EPIC images of the field. 

XMM-Newton's angular resolution does not resolve the pulsar from a nearby source of nearly equal (in CXO images) strengths rendering EPIC data useless for the pulsar spectral analysis. On the other hand, the EPIC background is higher than that of ACIS precluding the detection of fainter extended PWN emission and severely hampering spectral quality for the region where extended PWN emission is seen in the CXO images (see polygon region in bottom panel of Figure \ref{fig:cxo:spec:reg}). We attempted to extract EPIC spectra from a comparable in size circular region (after excluding the pulsar and a nearby point source). We only obtained a crude measurement of $\Gamma=1.6^{+0.8}_{-0.7}$ with $n_H$ being poorly constrained and dependent on the choice of background region with best fit values varying between 0.9 and $1.9\times10^{22}$ cm$^{-2}$ and large uncertainties. These values are compatible, albeit less accurate, with Chandra measurements (see Table \ref{tab:fits}).

\subsection{Transient Source CXOU J130131.0-625919}

\begin{figure}
\centering
\includegraphics[width=0.49\textwidth]{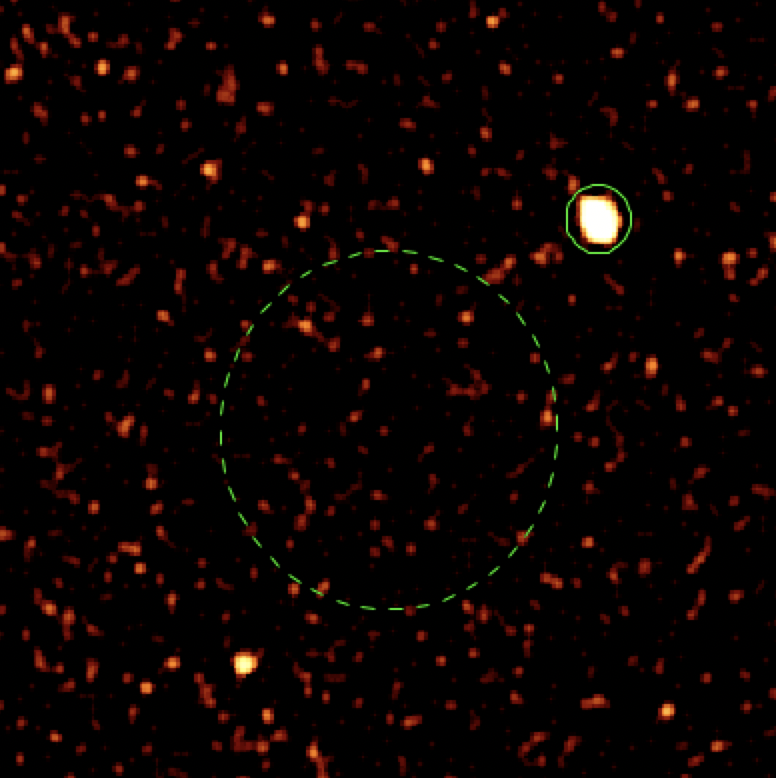}
\caption{Chandra image of the nearby variable source. Shown is obsID: 29080, which is when the large "flare" occurs. The region used for extracting the spectrum and light curve is shown in solid green and the associated background is indicated by the dashed line.}
\label{fig:var_src}
\end{figure}

\begin{figure}
\centering
\includegraphics[width=0.49\textwidth]{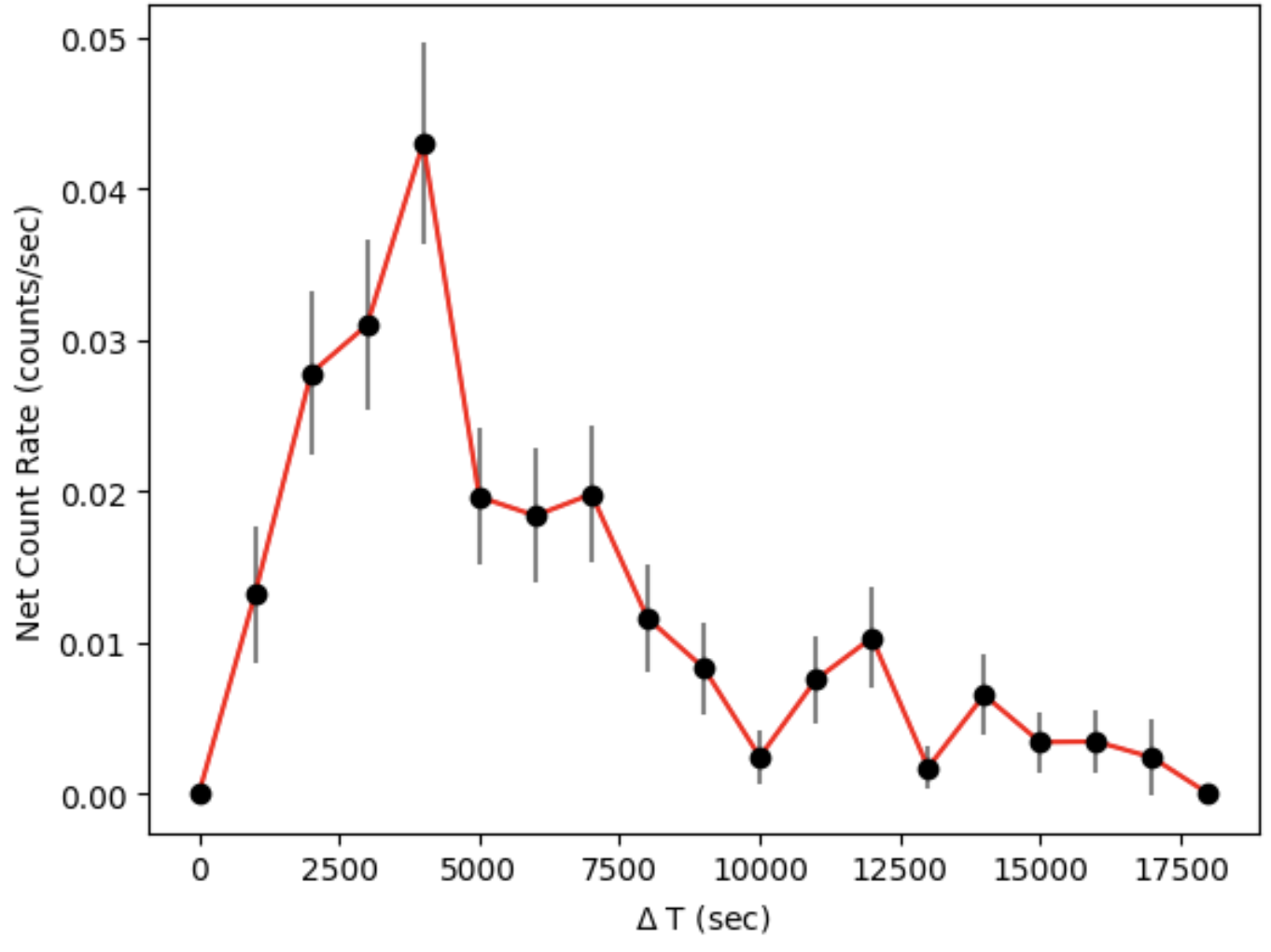}
\includegraphics[width=0.49\textwidth]{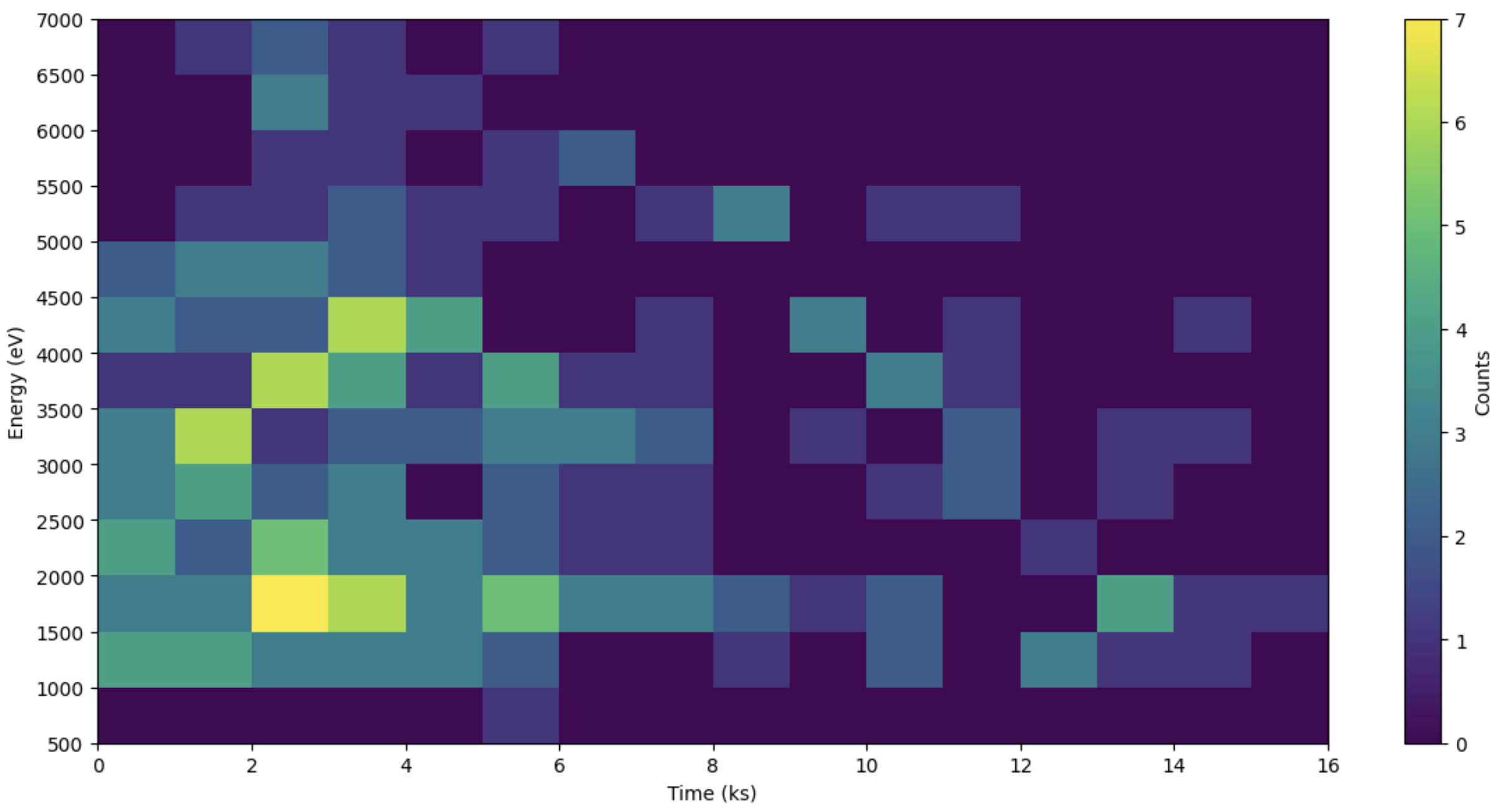}
\caption{Lightcurve and associated heatmap of the flaring source in obsID 29080.}
\label{fig:var_src_lc}
\end{figure}

Thanks to multiple CXO observations and also XMM-Newton observations of this field we discovered a flare from a transient  X-ray source that appears only in Chandra ObsID 29080  (15.87 ks exposure). 
To determine the source position the image was binned by a factor of four and the CIAO tool {\tt wavdetect} was used. The center of the source was found to be (RA, Decl.) = ($13^{\rm h}01^{\rm m}31.08^{\rm s}\pm 0.03^{\rm s},$ $-62^{\rm h}59^{\rm m}19.7^{\rm s}\pm 0.2^{\rm s}$).
These uncertainties do not include the astrometric error from Chandra's pointing uncertainty which is  $\sim1.17''$ (90\% confidence).\footnote{\url{https://cxc.harvard.edu/cal/ASPECT/celmon/}, accessed October 2026.}
The total number of detected source counts is 225 (99\% of which are due to the source). The source is imaged  $\sim6.5'$ off-axis on the ACIS-S6 chip (see Figure \ref{fig:var_src}). Thanks to its off-axis placement the pile-up is negligible even at the peak of the flare (see Figure \ref{fig:var_src_lc}).   
The lightcurve shown in the top panel was extracted 
following the CIAO thread\footnote{https://cxc.cfa.harvard.edu/ciao/threads/lightcurve/}.
The bottom panel,
created using the ciao task {\tt dmcopy} and binning events from the source region into energy bins of width 500 eV from 500 to 7000 eV and time bins of 1000 s, shows energy-resolved lightcurves as a heatmap. 
The spectrum of this source was extracted from the region shown in Figure \ref{fig:var_src}. 
XSPEC's implementation of the cstat statistic was used for the fit and counts were grouped
to have at least one count per bin prior to fitting. The energy was restricted to 0.5--7 keV. Fitting with a PL model modified by absorption ({\tt tbabs}) 
resulted in 
$\nh$ = $(0.6\pm0.5)\times10^{22}$ cm$^{-2}$, $\Gamma=1.4\pm0.3$, and $\mathcal{N} = (5.0\pm2.0)\times10^{-5}$ photons keV$^{-1}$ cm$^{-2}$ s$^{-1}$ at 1 keV with cstat/dof=114.69/167. The corresponding unabsorbed flux is $F_{0.5-7}=(3.10^{+0.09}_{-0.51})\times10^{-13}$ ergs cm$^{-2}$ s$^{-1}$,
and, assuming the same spectral shape throughout,
the peak flux was 
$F_{0.5-7}=(9.60^{+0.28}_{-1.58})\times10^{-13}$ ergs cm$^{-2}$ s$^{-1}$.

We also attempted to fit this source with an absorbed {\tt mekal} model.
Although the fit quality is acceptable (cstat/dof = 113.97/167)
the  resulting temperature can be arbitrarily high with a 3$\sigma$ lower limit of $\approx 8$ keV and best-fit value of $\approx 23$ keV. The best-fit $\nh=(0.6\pm0.3)\times10^{22}$ cm$^{-2}$ while the best-fit normalization corresponds to an emission measure of $\approx3\times 10^{55}$ cm$^{-3}$ at fiducial distance $d=3$ kpc.

This X-ray source has a plausible NIR counterpart 2MASS 13013105-6259198 located only $0.27''$ away (see Figure \ref{fig:transient:nir}). We retrieved all available optical-IR photometry from Vizier and fitted it with a set of PARSEC 1.2S + COLIBRI stellar evolution models\footnote{PAdova TRieste Stellar Evolution Code, see \citealt{2012MNRAS.427..127B}.} with solar-like abundances using spectra from Castelli–Kurucz atmosphere models \citep{castelli2004}. The best fit corresponds to a pre-main sequence star  ($\lesssim 1$ Myrs) with $A_V=5.1\pm0.5$,  $d=2-3.7 $,  $\log g=3.7$, and $T\approx5,000$ K. The fit, however, shows systematic residuals with data exceeding the fit at $\lesssim 600$ nm and in the reddest {\em Spitzer} IRAC band. The best-fit $A_V$ corresponds to $d=4.9-5.4$  kpc from DECaPS  3D dust map \citep{2025ApJ...992...39Z}, which is noticeably higher than that inferred from the stellar model fit.  We also inspected IRAC images which  show diffuse emission likely associated with star-forming activity mostly concentrated north of the Chandra transient (see Figure \ref{fig:transient:nir}).  We not that DECaPS 3D resolution is $1'$ while pockets of star-formation seen in IRAC images appear to be much more compact than that. This may explain discrepancy between the two distance estimates. Therefore, we below  will use  $d=3$ kpc. We also note that the best-fit $N_H$ implies (following relation of \citealt{2009MNRAS.400.2050G}) a factor of 2 lower $A_V$ than that inferred from the optical-NIR SED fitting.  

\section{Discussion} \label{sec:discussion}

\begin{figure}
\centering
\includegraphics[width=0.49\textwidth]{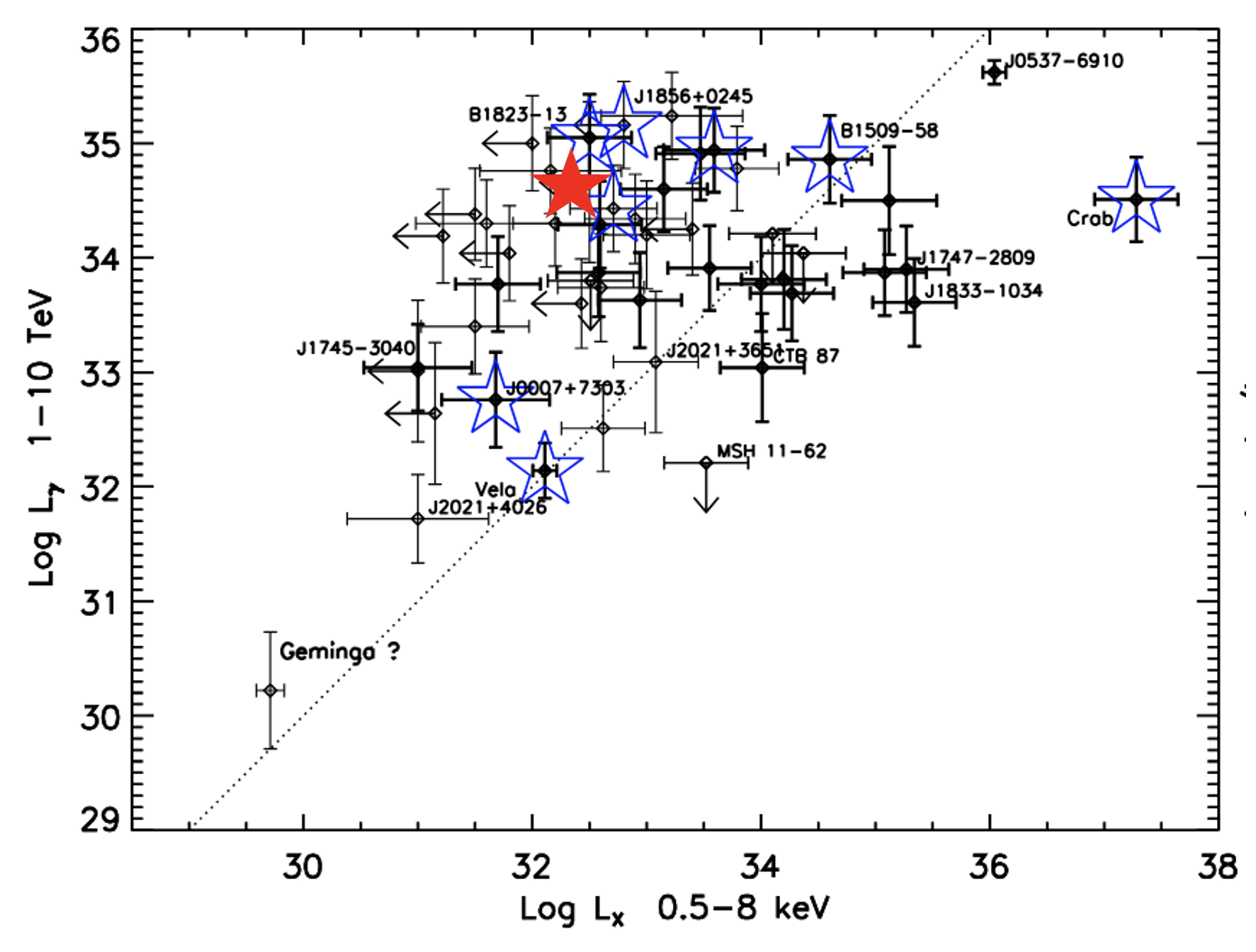}
\includegraphics[width=0.49\textwidth]{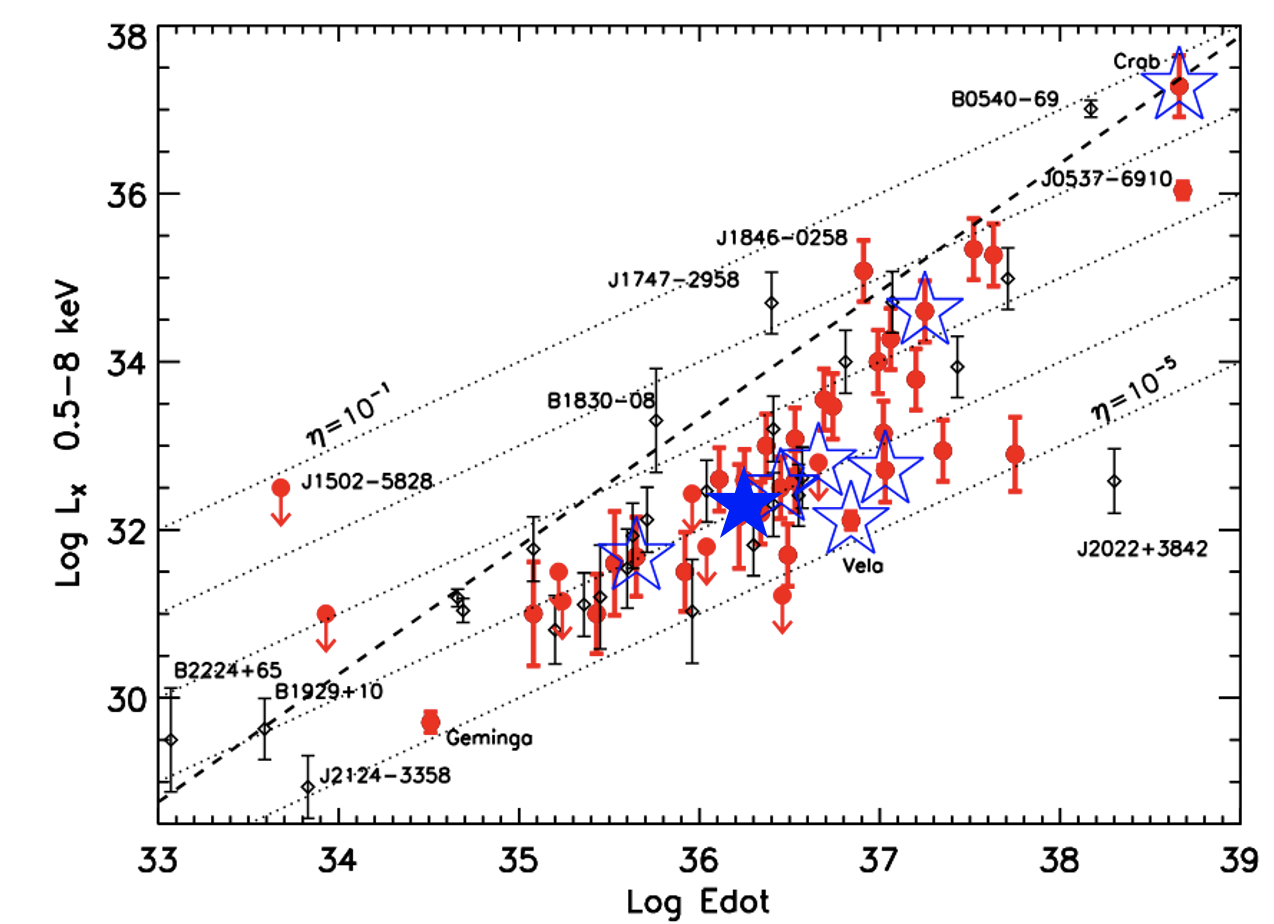}
\caption{Populations plots from \cite{2013uean.book..359K}
with \psr\ over-plotted with a red and blue star respectively.}
\label{fig:pop_plots}
\end{figure}

\begin{figure*}
\centering
\includegraphics[width=0.99\textwidth]{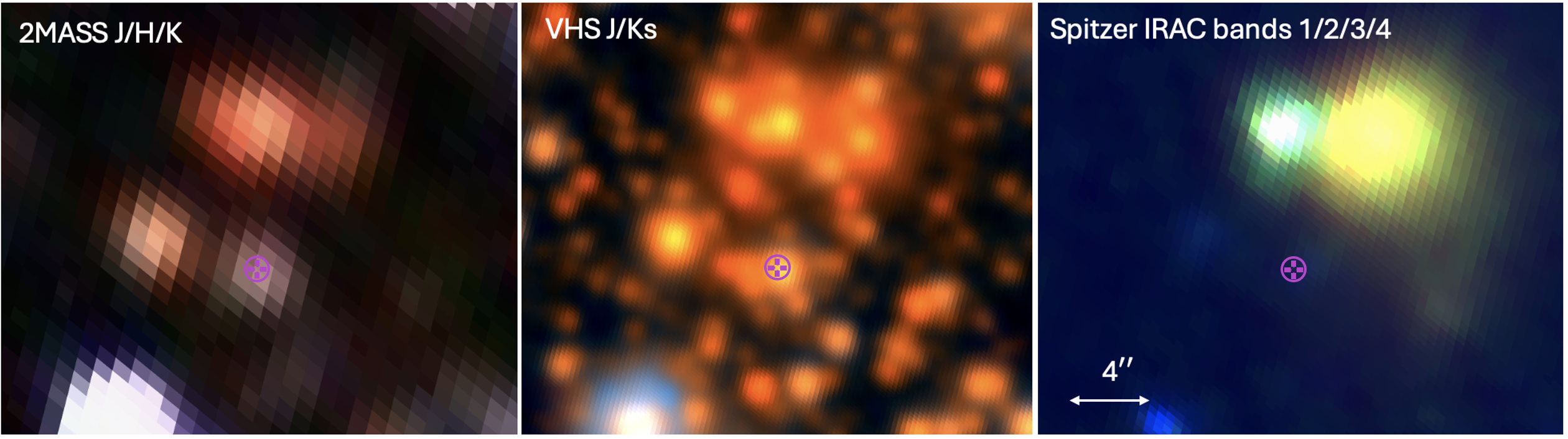}
\caption{Near-IR and IR images of  the X-ray transient source (shown by purple cross-hair, the purple circle shows the positional uncertainty estimated from {\tt wavdetect} and Chandra's pointing uncertainty) region obtained via  EsaSky. 2MASS 13013105-6259198 is closest to the cross-hair (centered at the Chandra transient source position) near the center of the images. }
\label{fig:transient:nir}
\end{figure*}

An evolved, synchrotron-underluminous PWN scenario for \hess\  has been previously  suggested by \citep{hess2012}. 
The small ($\lesssim 1.5'$) and faint X-ray nebula near J1301 shows some extension to the east (toward the center of HESS J1303-631; see Figure \ref{fig:xmm:mw}), supporting the association.  However, this factor alone, can hardly be taken as decisive evidence of an association with  \hess. The stronger support for the association comes from the hard, extended GeV counterpart \citep{zhou2023} and the energy-dependent TeV
morphology \citep{hess2012} which shows spectral hardening toward \psr. The compact X-ray PWN radiative efficiency, $L_X/\dot{E}\approx10^{-4}$, at d=4.7 kpc, is comparable to  those of many other pulsars with similar $\dot{E}$. On the other hand, the TeV luminosity is relatively high compared to other PWNe detcted in TeV (see Figure \ref{fig:pop_plots}).

The shape of the extended X-ray PWN component is surprisingly anisotropic and could be interpreted as a tail behind the pulsar moving in the western direction. However, the PWN emission is far too faint to say this with certainty. This example 
demonstrates the difficulty in understanding  
the nature of TeV sources
and identifying their X-ray counterparts, particularly in those cases when the existence  of a young radio pulsar have not yet been established 
through the detection of pulsations. Deep radio observations from modern surveys sensitive to extended structures (such as MeerKAT GPS) often show some extended emission, which, however, may not correlate well with TeV or X-ray emission. This 
seems to be the case for \psr\ and \hess\  (see Figure \ref{fig:xmm:mw}). Although the extended radio emission is clearly seen around \psr\ on $\sim$arcminute scales, it mostly extends west of the pulsar. It also shows a complex structure which could be attributed to the interaction with the reverse shock of the host SNR. However, unlike Vela X where both radio and TeV emission appear to trace the  relic PWN displaced by the reverse shock,  the TeV emission of \hess\ is  displaced in the opposite direction from the radio emission. It may be that the arcmin-scale radio emission is not related to PWN. Polarization and spectral measurements  (which is beyond the scope of this study) can help to establish its origin. There also appears to be a relatively bright, unresolved source imaged in  MeerKAT GPS coincident with \psr. Alternatively, it could be that enough time has passed for the most energetic (TeV emitting) electrons to diffuse out to much larger distances than the size of the PWN swept and compressed by the reverse shock.   

Other studies have faced similar challenges. 
For example, \cite{2009ApJ...690..891K} and \cite{2017ApJ...841...81H} 
analyzed a deep 60 ks Chandra observation of the young pulsar PSR J1617-5055, offset by about 10$'$ from the TeV source HESS J1616-508, to explore a 
possible association. Faint, 
diffuse emission surrounding the pulsar was resolved by Chandra with no preferential extension towards the TeV source center.
It corresponded to a low radiative efficiency $L_X/\dot{E}\sim10^{-4}$.  The authors could not confirm the association from the X-ray data alone, and archival radio/IR imaging indicated that the TeV emission was likely also powered, at least in part, by an unrelated SNR or star-forming region.

The transient source's most likely nature appears to be a pre-main sequence star (PMS) flare, given its location near/within the compact region of star-forming activity seen in IRAC images and the young age inferred from optical-IR SED fitting. The duration of the flare is compatible with those of PMS stars which typically follow the fast-rise-slow-decay profile (see Figure \ref{fig:var_src_lc}). However, the X-ray luminosity of $L_{X}=3.34\times10^{32}$ erg s$^{-1}$
($L_{X, {\,\rm peak}}=10.3\times10^{32}$ erg s$^{-1}$), in 0.5-7 keV  for a fiducial distance of 3 kpc and the high temperature $\gtrsim8$ keV, are rather extreme even for a young star (cf.\ Figure 9 of \citealt{2025ApJ...980...57G}). The inferred emission measure is compatible with those  PMS star flares but too large for main sequence star flares. An alternative interpretation could be an active binary (e.g., RS CVn) with young stars. It could also be another type of binary with a compact object, if the 2MASS source is a spurious association.
The large offset between the TeV source and the transient makes it unlikely to be related to the TeV source under any of the above scenarios. 

\section{Summary} \label{sec:summary}

\begin{enumerate}

\item We present new, deep joint \textit{Chandra} and \textit{XMM-Newton}
observations of the X-ray counterpart to the PWN HESS~J1303$-$631, likely
powered by PSR~J1301$-$6305. The \textit{Chandra} ACIS-S imaging, totaling
108\,ks, resolves the field into a compact pulsar source, a faint compact
nebula, and an extended diffuse component, successfully separating these
from a nearby, comparably bright confusing source that remains blended even
in the new, deep \textit{XMM-Newton} data -- underscoring the need for
\textit{Chandra}'s angular resolution in this crowded field.

\item Joint spectral fitting of the \textit{Chandra} data with an absorbed
power-law model yields a pulsar photon index of $\Gamma=2.02\pm1.04$ and an
unabsorbed 2--10\,keV flux of
$F_{2-10}=(1.30^{+0.20}_{-0.98})\times10^{-14}\,{\rm erg\,s^{-1}\,cm^{-2}}$,
consistent with typical values for pulsars powering PWNe. The compact
nebula spectrum is somewhat softer ($\Gamma=2.98\pm0.53$), while the
extended nebula component, with the compact nebula excised, is harder again
($\Gamma=1.20\pm1.25$); all three regions have comparable luminosities of a
few $\times10^{31}\,{\rm erg\,s^{-1}}$ at $d=4.7$\,kpc.

\item Combined with the previously known hard, extended GeV counterpart
(4FGL~J1303.0$-$6312e), the energy-dependent TeV morphology -- a broader,
displaced nebula at $0.84$--$2$\,TeV that becomes more compact and closer to
the pulsar above 10\,TeV -- and the absence of a bright radio PWN, these
X-ray results reinforce the picture of HESS~J1303$-$631 as an evolved,
synchrotron-underluminous PWN whose X-ray nebula is far more compact than
its TeV nebula, as expected from the rapid cooling of the highest-energy
synchrotron-emitting particles.

\item Deep ATCA and SARAO MeerKAT GPS radio imaging reveal a shell-like SNR
candidate $15'$ east of the pulsar and arcminute-scale diffuse structure
extending mostly west of it, but neither correlates well with the compact
X-ray/TeV nebula. Unlike Vela X, where the radio and TeV emission trace a
common relic nebula displaced by the reverse shock, the radio and TeV
structures here are offset in opposite directions from the pulsar,
suggesting the extended radio emission may be unrelated to the PWN.

\item We serendipitously discovered a transient X-ray source,
CXOU~J130131.0$-$625919, in one of nine new \textit{Chandra} pointings. Its
hard, fading spectrum ($\Gamma=1.4\pm0.3$) and luminosity of
$\approx3\times10^{32}\,{\rm erg\,s^{-1}}$ at a fiducial distance of 3\,kpc,
together with a plausible young stellar near-infrared counterpart, most
closely match a pre-main-sequence star flare, though an active binary or a
compact-object binary cannot be excluded. Its large angular offset from
HESS~J1303$-$631 rules out a physical association with the TeV source.

\item The anisotropic, tail-like shape of the extended X-ray nebula and its
possible relation to the host SNR's reverse shock, along with the mismatch
between the X-ray/TeV and radio morphologies, remain open questions. Deeper
radio polarization and spectral-index mapping, together with a search for
X-ray pulsations, would help establish the physical relationship between
these components.

\end{enumerate}

\begin{acknowledgments}
We thank Kaleb Martin for his early exploration of this data during a NASA summer internship.

Support for this work
was provided by the National Aeronautics and Space
Administration through Chandra Award Number GO4-25039A issued by the Chandra X-ray Center, which is
operated by the Smithsonian Astrophysical Observatory
under contract NAS8-03060. J.H. acknowledges support from NASA under award
number 80GSFC24M0006.

This research employs a list of Chandra datasets, obtained by the Chandra X-ray Observatory, contained in~\dataset[DOI: 10.25574]{https://doi.org/10.25574/cdc.710}.

\end{acknowledgments}

\bibliographystyle{aasjournal}
\bibliography{refs}

\end{document}